\documentclass[twoside,12pt]{article}

\usepackage{amssymb}

\newcounter{muni}

\newcommand{\nc}{\newcommand}

\nc{\nf}{\infty}
\nc{\beq}{\begin{equation}}     \nc{\nnb}{\nonumber}
\nc{\eeq}{\end{equation}}       \nc{\beqa}{\begin{eqnarray}}
\nc{\dst}{\displaystyle}        \nc{\eeqa}{\end{eqnarray}}

\nc{\mc}{\mathcal}              \nc{\dg}{\dagger}       \nc{\si}{\sigma}
\nc{\bs}{\backslash}            \nc{\nl}{\newline}      \nc{\mb}{\mathbb}
\nc{\qq}{\quad\quad}            \nc{\ol}{\overline}
\nc{\pt}{\partial}
\nc{\alf}{\alpha}               \nc{\be}{\beta}         \nc{\ga}{\gamma}
\nc{\de}{\delta}                \nc{\eps}{\epsilon}
\nc{\vtht}{\vartheta}
\nc{\om}{\omega}                \nc{\vp}{\varphi}
\nc{\vsi}{\varsigma}
\nc{\vrho}{\varrho}             \nc{\tht}{\theta}       \nc{\la}{\lambda}
\nc{\Om}{\Omega}                \nc{\Ga}{\Gamma}        \nc{\De}{\Delta}
\nc{\Log}{{\rm Log }}           \nc{\tg}{{\rm tg }}     \nc{\tr}{{\rm tr
}}
\nc{\sh}{{\rm sh }}             \nc{\ch}{{\rm ch }}

\title{\bf{Dualised $\si$-models at the two-loop order}}
\author{Guy Bonneau and Pierre-Yves Casteill \thanks {\noindent
Laboratoire de Physique Th\'eorique et des Hautes Energies,
 Unit\'e associ\'ee au CNRS UMR 7589, Universit\'e Paris 7,
 2 Place Jussieu, 75251 Paris Cedex 05. casteill@lpthe.jussieu.fr}}

\begin{document}
\date{\today}
\maketitle

\begin{abstract} We adress ourselves the question of the quantum
equivalence of non abelian dualised $\si$-models on the simple example of the
T-dualised
$SU(2)\,\si$-model. This theory is classically canonically equivalent to
the standard chiral $SU(2)\,\si$-model. It is known that the equivalence
also holds at the first order in perturbations with the same $\be$
functions. However, this model has been claimed to be non-renormalisable
at the two-loop order. The aim of the present work is the proof that it
is - at least up to this order - still possible to define a correct
quantum theory. Its target space metric being only modified in a finite
manner, all divergences are reabsorbed into coupling and  fields
(infinite) renormalisations.
\end{abstract}



\vfill {\bf PAR/LPTHE/01-14}\hfill  March 2001
\newpage

\section{Introduction} The subject of classical versus quantum
equivalence of T-dualised $\si$-models  has been strongly studied in
recent years, and extensive reviews covering  abelian, non-abelian
dualities and their applications to string theory and  statistical
physics are available
\cite{{aab1},{aal3},{gpr}}. More  recent developements on the
geometrical aspects of duality can be found  in
\cite{Al1}.

The interpretation of T-duality as a canonical transformation,  for
constant backgrounds, was first given by \cite{{grav},{CZ}}. Its more general
formulation
\cite{aal2} was applied to the non-abelian case in
\cite{{Lo},{Sf}}.

After the settling of the classical equivalence,  the most interesting
problem was its study at the quantum level.  This was done mostly for
dualisations of Lie groups, with emphasis put  on $\,SU(2).$ For this
model the one-loop equivalence was established  in \cite{{fj},{ft}}. This
one-loop quantum equivalence was recently settled for the  general class
of models built on
$G_L\times G_R/G_D,$ with an arbitrary breaking of $G_R$ \cite{cv}. An
interesting intermediary result is an expression for the Ricci tensor of
the dualised geometry (with torsion)  exhibiting its dependence with
respect to the geometrical  quantities of the original model. In the
same work, the two-loop renormalisability problem was  tackled and the
need for extra (non-minimal) one-loop order finite counter-terms was
emphasized. Some years ago, it was noted that in the minimal dimensional scheme,
 two-loop renormalisability does not hold for the SU(2) T-dualised model \cite{st,kt} .

The aim of the present work is a more precise analysis of this two-loop
(in)equivalence for the non-abelian T-duality\footnote{For abelian T-duality,
 similar works were achieved in \cite{kt2,km}.}, still on the simple example of the original SU(2) T-dualised
model.

The main remark is that, part of the isometries being somehow lost,
the T-dualised models are not - as they should be if one wants to give an
all-order analysis - defined by a sufficent system of Ward identities.
For example, in our simple case there is, {\it a priori } only a linear
SU(2) [or O(3)] invariance, and any O(3) invariant action is allowed (
let us remind the reader that in higher-loop corrections to a classical
action, all the terms which are not prohibited by some reason such as
power counting, isometries or conservation laws..., would appear). To
our present knowledge, the extra constraints coming from the origin of
the model (dualisation of an
$(SU(2)_L\times SU(2)_R)/SU(2)_D$ chiral model) are not understood
\footnote{In \cite{bd} the quantisation of  a U(1)-invariant non-linear
$\sigma$ model, the so-called Complex SineGordon model, was performed by
imposing as extra constraints its classical property of factorisation and
non-production ; there it was shown that {\bf definite extra finite}
one-loop counter-terms are needed to enforce this property to one-loop
order and then they also restore the two-loop renormalisability.} . As
it is highly probable that they are linked with the space-time dimension,
 it is not surprising that a minimal dimensional
renormalisation scheme fails : as is well known, when the regularization
method does not respect all the properties that define the theory,
extra finite counter-terms are needed
\cite{bo}.

The content of this article is the following : in  Section 2 we recall
the expression of the classical action of the dualised theory and set the
notations.  In Section 3, we start from the corresponding  {\it a priori}
quantum  bare action and obtain through
$\hbar$ expansion the possible counter-terms that may be added to the
classical action in order to reabsorb the divergences. Then in Section 4
we give the 2-loop  divergences and in Section 5 we discuss how they
match with the  candidates in Section 3. Our result is that coupling
constant and field renormalisations (infinite and finite ones ) are not
sufficient to ensure the two-loop existence of the T-dualised theory but
the metric itself has to be deformed (in a {\bf finite way}). Some
concluding remarks are offered in Section 6.

\newpage

\section{ The classical action}

At the classical level and in light-cone co-ordinates, the dual action
can be writen \cite{{fj},{cv}} :
$$ S= \frac{1}{\la}\int G_{ij}\pt_{+}\phi^{i}\pt_{-}\phi^{j} \,,$$ where
$g_{ij}=G_{(ij)}$ is the target space metric and
$h_{ij}=G_{[ij]}$ is the torsion potential. The torsion $T_{ijk}$ is
defined by
$T_{ijk}=\frac{3}{2}\pt _{[i}h_{jk]}$. The connections with torsion
$\Ga_{\:jk}^{i}$ and without torsion $\ga_{\:jk}^{i}$ respectively write
:
$$\Ga_{\:jk}^{i}=\frac{1}{2}g^{is}(\pt_{j}G_{ks} +
\pt_{k}G_{sj}-\pt_{s}G_{kj}) =\ga _{\:jk}^{i} + T_{\:jk}^{i}
\,,\qq \ga_{\:jk}^{i}=\frac{1}{2}g^{is}(\pt_{j}g_{sk} +
\pt_{k}g_{sj}-\pt_{s} g_{jk})\,,$$ and the coresponding covariant
derivatives are :
\[
D_{i}k_{j}=\pt_{i}k_{j}-\Ga_{\:ij}^{s}k_{s}=\nabla_{i}k_{j}-T_{\:ij}^{s}
v_{s}\,,\qq D_{i}k^{j}=\pt_{i}k^{j}+\Ga_{\:is}^{j}k^{s}=\nabla_{i}k^{j}
+T_{\:is}^{j}v^{s}.
\] The Riemann tensor without torsion will be noted $R_{ij,kl}$ whereas
we will denote the one with torsion as ${\bar{R}}_{ij,kl}\,.$

The expression of the dualised target space metric $G_{ij}$ as a
function of the original one is well known and in \cite{cv} the various
geometrical quantities (Ricci tensor,..) were also related. In the
special case considered here, where the original model is the
$\frac{SU(2)\times SU(2)}{SU(2)}$ non-linear $\si$ model, the metric
writes :
\beq\label{metric} G_{ij}[\vec{\phi}\,] =
\frac{1}{1+\vec{\phi}^2}\left[\delta_{ij} + \phi^i\phi^j +
\epsilon_{ijk}\phi^k \right]\,,\eeq

\noindent where $\vec{\phi}$ is a SU(2) (real) vector representation and
the
$\phi^i\
\,,\ i=1,\,2,\,3,$ are the co-ordinates on the dualised manifold. Then
$\vec{\phi}^2$ is a
$SO(3)$ invariant and the symmetry is linearly realised. Torsion breaks
parity, but the model is invariant under the simultaneous change $\phi
\rightarrow -\phi$ and
$\epsilon_{ijk}
\rightarrow -\epsilon_{ijk}\,.$ Let us emphasize that no other local
symmetry exists for that model.

\section{ The two-loop order bare action}

In order to analyse the two-loop renormalisability of the dualised
$SU2\,\,\si$-model, we first examine all the possible ways to reabsorb
the divergences through local counter-terms. As usual, we allow  for
finite and infinite renormalisations of both fields and coupling. But,
as we shall see later on, this appears as insufficient  to reabsorb the
various divergences. Thus, we also allow for a finite deformation of the
classical metric and torsion potential
$g_{ij}+h_{ij}=G_{ij}$ to describe its quantum extension : of course,
this {\it \`a la Friedan} \cite{Fr} extension of the notion of
renormalisability involves {\it a priori} an infinite number of new
parameters. Let us emphasize that we shall consider {\bf only finite
deformations}.

Even if by doing so we obviously introduce too many parameters, we first
let them all independent in order to show the announced need for such
intrinsic metric deformation.

Let us first write the bare action :
\beq\label{action_nue} S^{o}= \frac{1}{\la^{o}}\int
G_{ij}^{o}\pt_{+}\phi^{oi}\pt_{-}
\phi^{oj}
\eeq  where :
\beq
\left\{
\begin{array} [c]{l}
\dst\frac{1}{\la^{o}}=\frac{1}{\la}\left[  1+\frac{\hbar\la}{2\pi}\left(
\frac{\Lambda_{1} }{\varepsilon}+b\right) +\left
(\frac{\hbar\la}{2\pi}\right)^2\left(\frac{c}{\varepsilon^{2}}
+\frac{\Lambda_{2}}{\varepsilon}+d\right) + \cdots \right] ,  \\[5mm]
\dst\vec{\phi}^{o}=\vec{\phi}+\frac{\hbar\la}{2\pi}\left(
\frac{\vec{v}_{1}(\vec\phi)}{\varepsilon }+\vec{w}_{1}(\vec\phi)\right) +
\left(\frac{\hbar\la}{2\pi}\right)^2\left(
\frac{\vec{v}_{2}(\vec\phi)}{\varepsilon^{2}
}+\frac{\vec{w}_{2}(\vec\phi)}{\varepsilon}+\vec{x}(\vec\phi)\right) +
\cdots ,\\[5mm]
\dst G_{ij}^{o}=G_{ij}+\frac{\hbar\la}{2\pi}\,\,\widetilde{G}_{ij}
+\left(\frac{\hbar\la}{2\pi}\right)^2{\hat G}_{ij} + \cdots
\end{array}
\right. \label{Z0}
\eeq

 \noindent To express (\ref{action_nue}) we shall need the Lie derivative
$\dst
\mathop{\cal L}_{\vec{k}}$ and a ``second order'' Lie derivative $\dst
\mathop{{\cal L}^{(2)}}_{\vec{k}}\,.$  Indeed, for any
 tensor $S_{ij}$ defined on a manifold with co-ordinates
$\phi^{j}\,,$ in a change of co-ordinates~:
$$S_{ij}^0(\vec{\phi^0})\pt_{+}\phi^{oi}\pt_{-}
\phi^{oj} = S_{ij}(\vec{\phi})\pt_{+}\phi^{i}\pt_{-}
\phi^{j}\,,$$ and if $\vec{\phi^0} = \vec{\phi} + \eta\vec{k}\,$ (note
that
$\vec{k}$ is not a vector field on the manifold) :
\beq\label{Liedef} S_{ij}(\vec{\phi}) = S_{ij}^0(\vec{\phi}) -
\eta\mathop{\cal L}_{\vec{k}}(S_{ij}^0(\vec{\phi})) +
\frac{1}{2}\eta^{2}\mathop{{\cal L}^{(2)}}_{\vec{k}}
(S_{ij}^0(\vec{\phi})) + {\cal O}(\eta^3).
\eeq We remind the reader that
\beq\label{Lie1}
\mathop{\cal L}_{\vec{k}}(S_{ij})=k^{s}\nabla_{s}S_{ij}+S_{sj}\nabla_{i}
k^{s} + S_{is}\nabla_{j}k^{s}.\eeq

\noindent One can show that :
\beq\label{Lie2}
\mathop{{\cal L}^{(2)}}_{\vec k\,\,\,\,\,\,\,\,}(S_{ij})=\mathop{\cal
L}_{\vec k}\left(  \mathop{\cal L}_{\vec k}(S_{ij})\right)  -\mathop{\cal
L}_{k^{s}\pt_{s}\vec k}(S_{ij})\ \,.\eeq

\noindent With $\nabla_i\,g_{jk} = 0\,,$ we rewrite the equations
(\ref{Lie1},\ref{Lie2}) for
$S_{ij}
\equiv G_{ij}$ as :
\beq
\left\{
\begin{array}[c]{l}
\dst\mathop{\cal L}_{\vec{k}}(G_{ij})=2D_{j}k_{i}+\pt_{[i}\zeta_{j]}
\qq ,\qq \zeta_{i} = 2 k^lG_{li}\\[5mm]
\dst\mathop{{\cal L}^{(2)}}_{\vec{k}\,\,\,\,\,\,\,\,}(G_{ij})=2k^{s}k^{u}
{\bar R}_{si,ju}+2D_{i}k^{s}D_{j}k_{s}-4T_{ius}k^uD_jk^s+\mathop{\cal
L}_{(k^{s}k^{u}\vec{\ga} _{su})}(G_{ij})+\pt_{[i}{\hat \zeta}_{j]} .
\end{array}
\right.\label{Lie}
\eeq
\noindent ${\hat \zeta}_i$ is some quantity whose computation is useless
as, in the same manner as ${ \zeta}_i\,,$ it gives a vanishing
contribution to the action or, the torsion potential being always
defined up to a gauge transformation, such term can always be put into
$h_{ij}$ (moreover, in our particular situation, the $O(3)$ symmetry
implies that such
$\pt_{[i}\zeta_{j]}$ terms vanish). Then, we shall not write them
anymore.

Then, expending (\ref{action_nue}) with the help of
(\ref{Z0},\ref{Liedef}), one gets the possible counter-terms at lowest
orders :

\begin{itemize}
\item   0 order in $\dst\frac{\hbar\la}{2\pi}$ :

$$\frac{1}{\lambda}G_{ij}\pt_{+}\phi^{i}\pt_{-}\phi^{j}$$

\item  first order in $\dst\frac{\hbar\la}{2\pi}$ :
\beq\label{1o}
\frac{1}{\la}\left[ \left( \frac{\Lambda_{1}}{\varepsilon} + b
\right)\,G_{ij}
   + \mathop{\cal L}_{\frac{\vec{v}_{1}}{\varepsilon} +
\vec{w}_{1}}(G_{ij})   + \widetilde{G}_{ij}  \right]
\pt _{+}\phi^{i}\pt_{-}\phi^{j}
\eeq

\item  at second order in $\dst\frac{\hbar\la}{2\pi}$ :
\begin{equation}
\begin{array} [c]{l}
\dst \frac{1}{\la}\left[  \frac{1}{\varepsilon^{2}}\left(
\cdots\right)  \right.
\\[5mm]
\dst + \left(  \frac{\Lambda_{1}}{\varepsilon}\,\widetilde{G}_{ij} +
\mathop{\cal L}_{\frac{\vec{v}_{1}}{\varepsilon}}(\widetilde{G}_{ij}) +
\frac{\Lambda_{1}}{\varepsilon}\mathop{\cal L}_{\vec{w}_{1}}(G_{ij}) +
b\mathop{\cal L}_{\frac{\vec {v}_{1}}{\varepsilon}}(G_{ij}) +
\frac{\Lambda_{2}}{\varepsilon}G_{ij} + \mathop{\cal L}_{\frac{\vec
{w}_{2}}{\varepsilon}}\,(G_{ij}) +\frac{1}{2}\left.
\mathop{{\cal
L}^{(2)}}_{\frac{\vec{v}_{1}}{\varepsilon}+\vec{w}_{1}}(G_{ij})\right|
_{\frac{1}{\varepsilon}}\right)  \\[5mm]
\dst\left.  + \left(  \cdots\right)  \right]  \pt_{+}\phi^{i}
\pt_{-}\phi^{j}
\end{array}
\label{2o1}
\end{equation}

\noindent where $\left.  \dst\mathop{Q}_{\:}^{\:}\right|
_{\frac{1}{\varepsilon}}$ means that we only take the term in
$\frac{1}{\varepsilon}$ in the expression $Q$.
\end{itemize}

\noindent As we don't consider the 3-loop order, in  expression
(\ref{2o1}) we only need the coefficient of
$\frac{1}{\varepsilon}$  (the double poles
$\frac{\hbar^2}{\varepsilon^2}$ are not new quantities as they are
directly related to first order simple poles and
 it has already been proved that the dualised
$SU2\,\si$-model is one-loop renormalisable \cite{fj}).

Using the following identity
between Lie derivatives :
$$\mathop{\cal L}_{\vec{X}}\mathop{\cal L}_{\vec{Y}} -
\mathop{\cal L}_{\vec{Y}}\mathop{\cal L}_{\vec{X}} = \mathop{\cal
L}_{\vec{Z}}\qq {\rm with}\ Z^i = X^j\partial_jY^i -
Y^j\partial_jX^i\,,$$ the term with the ``second
order'' Lie derivative may be re-expressed :

$$\varepsilon\left.\left[\mathop{{\cal L}^{(2)}}_{\frac
{\vec{v}_{1}}{\varepsilon}+\vec{w}_{1}}\,(G_{ij})\right|
_{\frac{1}{\varepsilon}}\right] = \mathop{{\cal
L}}_{\vec{v}_{1}} \mathop{{\cal
L}}_{\vec{w}_{1}}\,(G_{ij}) + \mathop{{\cal
L}}_{\vec{w}_{1}} \mathop{{\cal
L}}_{\vec{v}_{1}}\,(G_{ij}) -  \mathop{{\cal
L}}_{({v}_{1}^k\partial_k\vec{w}_{1} +
{w}_{1}^k\partial_k\vec{v}_{1})}(G_{ij}) =
2\left[\mathop{{\cal L}}_{\vec{v}_{1}} \mathop{{\cal
L}}_{\vec{w}_{1}}\,(G_{ij}) - \mathop{{\cal
L}}_{{v}_{1}^k\partial_k\vec{w}_{1}}\,(G_{ij})\right]\,.$$

\noindent So, the ${\cal O}(\hbar)$ term (\ref{1o}) may be rewriten as
:
$$ \frac{1}{\la}\left[
\frac{1}{\varepsilon}\left( \Lambda_{1}G_{ij} +  \mathop{\cal
L}_{\vec{v}_{1}}(G_{ij}) \right)\, +
\left(\mathop{\cal
L}_{\vec{w}_{1}}(G_{ij}) + b G_{ij} + \widetilde{G}_{ij}\right) \right]
\pt _{+}\phi^{i}\pt_{-}\phi^{j}\,,$$
and the ${\cal O}(\hbar)^2$ term (\ref{2o1}) as :
$$ \frac{1}{\la\varepsilon}\left[\Lambda_1\left(
\mathop{\cal
L}_{\vec{w}_{1}}(G_{ij}) + b G_{ij} +
\widetilde{G}_{ij}\right) + (\Lambda_{2} - b
\Lambda_{1}){G}_{ij} +
\mathop{\cal L}_{\vec{v}_{1}}\left(\mathop{\cal
L}_{\vec{w}_{1}}(G_{ij}) + b G_{ij} +
\widetilde{G}_{ij}\right) + \mathop{\cal L}_{(\vec{w}_{2} -
v_1^k\partial_k \vec{w}_{1})}(G_{ij}) \right]\,.$$

\noindent As a consequence, as expected, any term  $\dst \mathop{\cal
L}_{\vec{w}_{1}}(G_{ij}) + b G_{ij}$ may be reabsorbed into the finite
deformation
$\widetilde{G}_{ij}$ (and {\it vice-versa}) to the expense of a change in
the
${\cal{O}}(\hbar)^2 $ parameters : \beq\label{subs}
\widetilde{G}_{ij} + \mathop{\cal L}_{\vec{w}_{1}}\,(G_{ij}) + b
G_{ij} \rightarrow \bar{G}_{ij} \Rightarrow \left\{ \Lambda_2
\rightarrow \Lambda_2 -
b\Lambda_1
\,,\ \ {\vec{w_2}}
\rightarrow  {\vec{w_2}} - v_1^k\partial_k{\vec{w_1}} \right\}\,.\eeq

\vspace{1cm}

Finally, for the term in
$\frac1\varepsilon\left(\frac{\hbar\la}{2\pi}\right)^2$ in the bare
action, one has the following expression :

\beq
\frac{1}{\la}\left(\Lambda_{1}\widetilde{G}_{ij} +\mathop{\cal
L}_{\vec{v}_{1}}(\widetilde{G}_{ij}) + \Lambda_{2}G_{ij}+ \mathop{\cal
L}_{\vec{W}_2}(G_{ij}) + {\cal H}_{ij}(v_1,w_1)\right) ,
\label{2o}
\eeq where
\beq
\left\{\begin{array}[c]{l}
\vec{W}_2=\vec{w}_{2}+ b\,\vec{v}_1+\Lambda_{1}\vec{w}_{1} +
v_1^{s}w_1^{u}\vec{\ga}_{su}\\[2mm] {\cal H}_{ij}(v_1,w_1)=v_1^{s}w_1^{u}
{\bar R}_{is,uj}+D_{i}v_1^{s}D_{j}w_{1s}-2T_{ius}v_1^uD_jw_1^s+(\vec{v}_1
\leftrightarrow
\vec{w}_1) .
\end{array}\right. \label{21}
\eeq

\section{The two-loop order divergences}

We use the expression of the covariant divergences  given by Hull
and Townsend \cite{HT}
\footnote{\,We checked for our example that the two other calculations in
\cite{{mt},{z}} give the same result. } , in the
background field method and in the {\bf minimal} dimensional
scheme,  up to the two-loop order :

\beq\label{div}
\left\{
\begin{array} [c]{l}
\dst Div_{ij}^{1}=-\frac{\hbar}{2\pi\varepsilon}Ric_{\,ij}\\[3mm]
\dst Div_{ij}^{2}=-\frac{\hbar^{2}\la}{8\pi^{2}\varepsilon}\left(
\bar{R}_{i}
^{\;\,klm}(\bar{R}_{klmj}-\frac{1}{2}\bar{R}_{lmkj})+2T_{\:mn}^{k}T^{lmn}\bar
{R}_{kijl}\right)
\end{array}
\right.
\eeq

In order to ensure the renormalizability of the theory, these divergences
should match with the candidate counter-terms given by (\ref{1o}) and
(\ref{2o}) :

\beq\label{ct}
\left\{
\begin{array} [c]{l}
\dst CT_{ij}^{1}=\frac{\hbar}{ 2\pi\varepsilon}\left(
\Lambda_{1}G_{ij}  +\mathop{\cal L}_{\vec{v}_{1}}(G_{ij})\right) \\
\dst CT_{ij}^{2}=\frac{\hbar^{2}\la}{ 4\pi^2\varepsilon}\left(
\Lambda_{1}\widetilde{G}  _{ij}+\mathop{\cal
L}_{\vec{v}_{1}}(\widetilde{G}_{ij})+\Lambda_{2}  G_{ij}+\mathop{\cal
L}_{\vec{W}_{2}}(G_{ij})+{\cal H}_{ij}(v_1,w_1)\right)
\end{array}
\right.
\eeq It has been previously proven \cite{cv} that the dualised metric is
quasi-Einstein as soon as the original metric is Einstein. In our special
case, we get :
\beq\label{renor1} Ric_{\,ij}= \Lambda\,G_{ij} +
2D_{j}v_{i}\quad,\quad\Lambda = \Lambda_{1}=\frac{1}{2}\quad,\quad
\vec{v} = \vec{v}_{1} =\frac{1}{2}\left(
\frac{1-\phi^{2}}{1+\phi^{2}}\right)  \vec{\phi} \eeq

\noindent The addition to the effective action of a $\hbar$ finite
deformation of the metric and of some finite renormalisations for the
coupling and fields (non-minimal scheme) modifies the $\hbar^2$
divergences. The additional term is easily obtained as

$$-\frac{\hbar}{2\pi\varepsilon}\dst \left\{ Ric_{\,ij}\left(
G_{kl}+\frac{\hbar\la}{2\pi}(\mathop{\cal L}_{\vec{w}_{1}}(G_{kl} ) +
b\,G_{kl} + \widetilde{G}_{kl})\right) - Ric_{\,ij}(G_{kl}) \right\}
\equiv -
\frac{\hbar^{2}\la}{4\pi^2\varepsilon}\Delta_{ij}
+{\cal{O}}(\hbar^{3})\,.$$

Here also, only the combination $\dst\widetilde{G}_{ij} + \mathop{\cal
L}_{\vec{w}_{1}}(G_{ij}) + b G_{ij}$ appears. Then, we could decide to
reabsorb
$b\,G_{ij}$ and $\dst\mathop{\cal L}_{\vec{w}_{1}}(G_{ij})$ into
$\widetilde{G}_{ij}\,,$ but, as announced at the beginning of Section 3,
in order to see if they would be sufficient by themselves, we keep them
apart in a first step.

Finally, the dualised $SU2$
$\sigma$-model will be renormalisable at two loops if and only if we can
find  $\left\{ \widetilde {G}_{ij}[\,{\vec{\phi}}]\,, \ b\,,\
{\vec{w}_{1}}[\,{\vec{\phi}}]\,;\ \Lambda_{2}\,,\
\vec{W}_{2}[\,{\vec{\phi}}]\right\}$ such that :

\begin{equation} Div_{ij}^{2} -  \frac{\hbar^{2}\la}{4\pi^2\varepsilon}
\Delta_{ij} + \frac{\hbar^{2} \la }{ 4\pi^2\varepsilon}\left(
\Lambda_{1}\widetilde{G}_{ij}+\mathop{\cal L}_{\vec{v}_{1}
}(\widetilde{G}_{ij})+\Lambda_{2}G_{ij}+\mathop{\cal L}_{\vec{W}_{2}
}(G_{ij})+ {\cal H}_{ij}(v_1,w_1)  \right) = 0 \label{eq}
\end{equation}

\section{Results}

According to the linearly realised symmetry of the T-dualised
$SU2\,\si$-model, the finite deformation of the metric
 $\widetilde {G}_{ij}$ and the vectors $\vec{w}_{1}(\phi)$ and
$\vec{W}_{2}(\phi)$  respectively write :

$$
\widetilde{G}_{ij}=\alpha(\tau)\delta_{ij}+\beta(\tau)\phi^{i}\phi^{j}
+\epsilon_{ijk}\gamma(\tau)\phi^{k}\quad,
\quad\vec{w}_{1}=w_{1}(\tau)\vec{\phi}
\quad,\quad\vec{W}_{2}=W_{2}(\tau)\vec{\phi}$$
 where $\tau=\vec{\phi}^{2}$. Moreover, the symmetry also implies that
terms of the form
$\pt_{[i}k_{j]}$ or of the form $k^{s}K^{u}\ga^t_{su}$ are equal to
zero.  It is then possible to  re-express (\ref{eq}) as a set of three
linear differential equations~:

\beqa\label{eq1} W_2(\tau) & + & \frac{(1+\tau)\Lambda_2}{2} +\frac{45 +
68\tau -18\tau^2 -12\tau^3 -3\tau^4}{16{(1+\tau)}^3} -
\frac{1-\tau}{(1+\tau)^2}w_1(\tau) \nnb\\ & = &\frac{3+10\tau+5\tau^2
+2\tau^3}{4(1+\tau)}\alpha(\tau)
 +  \frac{4+5\tau +6\tau^2 +\tau^3}{4(1+\tau)}\beta(\tau)
-\frac{3(1+\tau)(3+\tau)}{2}\gamma(\tau) \\ & -
&\frac{4+11\tau+5\tau^2-\tau^3}{2}\alpha'(\tau) +
\frac{\tau}{2}\beta'(\tau) -\tau(1+\tau)(3+\tau)\gamma'(\tau)
-\tau{(1+\tau)}^2\alpha''(\tau)  \nnb
\eeqa

\beqa\label{eq2} 3\Lambda_2
-\frac{3(-5+60\tau+10\tau^2+12\tau^3+3\tau^4)}{8{(1+\tau)}^4} & = &
\frac{(7+10\tau)}{2}\alpha(\tau) +\frac{(12+5\tau)}{2}\beta(\tau)
-3(11+5\tau)\gamma(\tau) \nnb\\ + (-17-22\tau+9\tau^2)\alpha'(\tau) & + &
(5+4\tau+\tau^2)\beta'(\tau)  - 2(5+2\tau)(3+5\tau)\gamma'(\tau) \\ +
2(-5-19\tau-12\tau^2+\tau^3)\alpha''(\tau) + 2\tau\beta''(\tau) & - &
4\tau(1+\tau)(3+\tau)\gamma''(\tau) - 4\tau{(1+\tau)}^2\alpha^{(3)}(\tau)
\nnb
\eeqa

\beqa\label{eq3}
\Lambda_2 + \frac{3(1 -\tau)(13+6\tau+\tau^2)}{8{(1+\tau)}^3} & = &
\frac{(5+2\tau)}{2}\alpha(\tau) +\frac{(6+\tau)}{2}\beta(\tau)
-(17+3\tau)\gamma(\tau) \nnb\\ + (-7+\tau)(3+\tau)\alpha'(\tau)
-2(-5+6\tau+\tau^2)\gamma'(\tau) & - & 2\tau(3+\tau)\alpha''(\tau)
+4\tau\gamma''(\tau)
\eeqa

\noindent The need for a true deformation ${\widetilde G}_{ij}$
immediatly appears : setting both $\alpha(\tau)$, $\beta(\tau)$ and
$\gamma(\tau)$ to zero, equations (\ref{eq2}) and (\ref{eq3}) cannot be
satisfied, even if \footnote{\,One notices
also that  the parameters $b$ and $\vec{w}_1$ do not appear in
(\ref{eq2}) and (\ref{eq3}). So, the {\bf existence} of some solution to
this set of differential equations is independent of the finite
renormalisations of both coupling and fields, as is usual in
perturbation theory. This freedom corresponds to a change of
renormalisation scheme. This absence is only true if we take the very
vector
$\vec{v}_1$ (\ref{renor1}) that reabsorbs the divergences at the one-loop
order : otherwhise,
$\vec{w}_1({\vec{\phi}})$ would appear in (\ref{eq2}) and (\ref{eq3}).
This is a check of a correct renormalisation at the one-loop order.} we
allowed for some finite renormalisations of the coupling ($b$) and field
($\vec{w}_1({\vec{\phi}})$), both hidden into the vector
$\vec{W}_2({\vec{\phi}})$ (see equation (\ref{21})). Then, as first proven in \cite{st,kt}, we have checked
that :
\newline {\bf In a purely dimensional scheme (even with non minimal
subtractions), \\ the dualised $\mathbf {SU(2)\ \si}$ model is not
renormalisable at the two-loop order}.

\vspace{1cm} 

So, from the discussion in the previous
Sections, and without restricting the generality of our analysis, one
can take
$b$ and
$\vec{w}_1({\vec{\phi}})$ as vanishing quantities.

\vspace{0.5cm}

\noindent Remarks :
\begin{itemize}
\item As $\Lambda_2$ is not a function, but a constant,
differentiating equations (\ref{eq2}) and (\ref{eq3}) will relate
$\alpha(\tau)$,
$\beta(\tau)$ and
$\gamma(\tau)$. Then, as soon as ${\widetilde
G}_{ij}\,,$ the finite one loop renormalisation, has been definitely set, equation (\ref{eq1}) will 
give the infinite two-loop
renormalisations
$\vec{W}_2({\vec{\phi}})$ and $\Lambda_2\,.$ 

\item From the previous discussions, we know that ${\widetilde G}_{ij}$
will be fixed up to some $\dst{\check b}\,G_{ij}+\mathop{\cal
L}_{\vec{\check W}({\vec{\phi}})}(G_{ij})$; it is then natural to
use this freedom, for example to reabsorb $\alpha(\tau)\,,$ and to
redefine
${\widetilde G}_{ij}$ such that :
\beqa\label{irrelevant} {\widetilde G}_{ij}= {\check
b}\,G_{ij}+\mathop{\cal L}_{\vec{\check W}}({\vec{\phi}})(G_{ij}) +
{\check{G}}_{ij} & ,& {\vec{{\check W}}}({\vec{\phi}}) =
{\check W(\tau){\vec{\phi}}}\nnb\\ {\rm with}\ {\check W}(\tau) =
\frac{(1+\tau)^2}{2}\alpha(\tau) - \frac{{\check b}(1+\tau)}{2} &  \Rightarrow &
  {\check {G }}_{ij} =
{\check\beta}(\tau)\phi^{i}\phi^{j} +
\epsilon_{ijk}{\check\gamma}(\tau)\phi^{k}\nnb\\
{\rm with}\qq {\check\beta} = \beta -\frac{{\check b}}{1+\tau}
-\frac{2(2+\tau){\check W}}{(1+\tau)^2} -4{\check W}' & {\rm and}  &
{\check\gamma} =  \gamma  -\frac{{\check b}}{1+\tau}
-\frac{(3+\tau){\check W}}{(1+\tau)^2}\,.
\eeqa
\noindent We know that, when expressed as functions of
$\check\beta(\tau)$ and $\check\gamma(\tau)\,,$ equations
(\ref{eq1},\ref{eq2},\ref{eq3}) remain unchanged,  up to the substitutions
discussed in Section 3 [equation (\ref{subs})] :
\beqa\label{substitution}
\Lambda_2
\rightarrow {\check
\Lambda}_2 & = & \Lambda_2 + \frac{\check b}{2}\,,\nnb\\
W_2({\vec{\phi}})
\rightarrow {\check{W}}_2(\tau) & = & W_2(\tau) +
{\check b}\frac{1-\tau}{2(1+\tau)} + \frac{1}{2}{\check{W}}(\tau) +
\frac{1-\tau}{2(1+\tau)}[ {\check{W}}(\tau)
+2\tau {\check{W}}'(\tau)]\,.\eeqa
\end{itemize}

\noindent Equations (\ref{eq2}) and (\ref{eq3}) give ${\check
\beta}(\tau)$ as a function of ${\check
\gamma}(\tau)$ which itself satisfies a non-homogeneous linear fourth
order differential equation :

\beqa
 (a) & {\check\beta}(\tau) & -\ \frac{{\check b}
+2 \Lambda_2}{6+\tau}  =
\frac{3(1-\tau)(13+6\tau+\tau^2)}{4(6+\tau)(1+\tau)^3} +
\frac{2(17+3\tau)}{6+\tau}{\check\gamma}(\tau)
- \frac{4(5-6\tau-\tau^2)}{6+\tau}{\check\gamma}'(\tau) \nnb \\
& - & \frac{8\tau}{6+\tau}{\check\gamma}''(\tau)\nnb \\
 (b) & {\check\gamma}^{(4)}(\tau) & +
\ \frac{(6 -\tau)(7+\tau)}{\tau(6+\tau)}{\check\gamma}^{(3)}(\tau)  +
\frac{1260-276\tau-91\tau^2+3\tau^3+\tau^4}{4\tau^2(6+\tau)^2}
{\check\gamma}''(\tau) +  \nnb\\ & +
& \frac{-120+254\tau+57\tau^2+3\tau^3}
{8\tau^2(6+\tau)^2}{\check\gamma}'(\tau) -
\frac{138+25\tau+\tau^2}{8\tau^2(6+\tau)^2}[{\check\gamma}(\tau) -
\frac{{\check b} +2 \Lambda_2}{2}] = \nnb\\
& = & -\frac{3(6402-8681\tau-5856\tau^2-22\tau^3+390\tau^4+39\tau^5)}
{64\tau^2 (1+\tau)^5 (6+\tau)^2}\ . \label{eqfinale}
\eeqa

\noindent Note that under the change \beq\label{30}
\Ga(\tau) =
[{\check\gamma}(\tau) -
\frac{{\check b} +2 \Lambda_2}{2}]\,,\qq B(\tau) =
[{\check\beta}(\tau) -
3({\check b} +2 \Lambda_2)]\,,\eeq the parameter ${\check b}$ and the
constant $\Lambda_2$ disappear from the set (\ref{eqfinale}). Then,
$\Lambda_2$ being an unknown constant, the general solution of the
differential equation (\ref{eqfinale}-b) will be $${\check\gamma}(\tau)
= \Gamma(\tau) +c\,, {\rm where\ c\ is\ an\ arbitrary\ constant}\,,$$ and
the two-loop coupling constant renormalisation $\Lambda_2$ will be :
$$\Lambda_2 = c - \frac{\check b}{2}\,.$$

\noindent The  model will be renormalisable up to two loops iff equation
(\ref{eqfinale}-b) , where ${\check\gamma}(\tau)$ has been replaced by
$\Gamma(\tau)$ according to (\ref{30}), has a solution which is analytic
near $\tau = 0$. In order to reach such a conclusion, we use the method
of Frobenius for linear  differential equations \cite{Ince}. $\tau = 0$
is a regular singularity (notice that we are only interested in $\tau
\ge 0$). The indicial equation of the linear differential equation
(\ref{eqfinale}-b) around the singular point
$0$ has four different solutions : $\nu = -\frac{3}{2},\ -\frac{1}{2},\
0,\ 1\,.$ For each one, we can find convergent series
$\dst\tau^{\nu}\sum_{n =0}^{\infty}c_{n}\tau^{n}$ that are independent
solutions of the homogeneous equation associated to (\ref{eqfinale}-b).
We give here the first terms of such series (it happens that for
$\nu=-\frac{3}{2}$ we have an exact solution ) :

\beq
\begin{array}[l]{llllllll} {\check\gamma}_{-\frac{3}{2}}(\tau) &=&
\dst \frac{1}{\tau^{\frac{3}{2}}} +\frac{1}{20\sqrt{\tau}} -
\frac{\sqrt{\tau}}{20} &,& {\check\gamma}_{-\frac{1}{2}}(\tau) &=&
\dst\frac{1}{\sqrt{\tau}}\left(1-\frac{11}{6}\tau +
\frac{35}{108}\tau^2+\cdots \right)    \\[3mm] {\check\gamma}_0(\tau) &=&
\dst 1+\frac{23}{840}\tau^2 + \cdots  &,& {\check\gamma}_1(\tau) &=&
\dst\tau\left(1+\frac{1}{42}\tau -\frac{1}{324}\tau^2+
\cdots\right)
\end{array}
\eeq

\noindent Then, we use the method of variation of parameters to find
$\la_{-\frac{3}{2}}(\tau)\,,\ \la_{-\frac{1}{2}}(\tau)\,,\ \la_0(\tau)$
and
$\la_1(\tau)$ such that
$$\Gamma(\tau)
=\la_{-\frac{3}{2}}(\tau){\check\gamma}_{-\frac{3}{2}}(\tau) +
\la_{-\frac{1}{2}}(\tau){\check\gamma}_{-\frac{1}{2}}(\tau) +
\la_0(\tau){\check\gamma}_0(\tau) + \la_1(\tau){\check\gamma}_1(\tau) $$
\noindent is the general solution of the inhomogeneous  equation
(\ref{eqfinale}-b) where ${\check\gamma}(\tau)$ has been replaced by
$\Gamma(\tau)$ according to (\ref{30}).

The first terms in the expansion of these functions are :
\beq
\begin{array}[l]{ll}
\la_{-\frac{3}{2}}(\tau) =\dst \la_{-\frac{3}{2}}^o + \tau^{\frac{7}{2}}
\left(\frac{1067}{1680} -\frac{13691}{3780}\tau + \cdots\right) &,
\dst\la_{-\frac{1}{2}}(\tau) =\dst \la_{-\frac{1}{2}}^o +
\tau^{\frac{5}{2}}
\left(-\frac{1067}{240}+\frac{2543509}{100800}\tau +
\cdots\right)\nnb\\[2mm]

\dst\la_0(\tau) =\dst \la_0^o + \frac{1067}{192}\tau^2 -
\frac{9805}{288}\tau^3 + \cdots &,
\dst\la_1(\tau) =\dst \la_1^o - \frac{1067}{480}\tau +
\frac{27887}{5760}\tau^2 +\cdots \nnb
\end{array}
\eeq
 The analyticity requirement near
$\tau = 0$ enforces the choice
$\la_{-\frac{3}{2}}^o = \la_{-\frac{1}{2}}^o = 0 \,;$
${\check\gamma}(\tau)$ is then expressed as a convergent series in
$\tau\,,$ and the same will be true for ${\check\beta}(\tau)\,.$ The
final expression for the deformation $\widetilde{G}_{ij}$ depends on 3
constants [ $c\,,\ \la_0^o$ and $\la_1^o$] and an arbitrary function
[${\check W}(\tau)$] and is given by the three functions
: \beqa\label{resultat}
& \alpha(\tau) & =  \frac{{\check b}}{1+\tau} + \frac{2{\check
W}}{(1+\tau)^2}\,,\nnb\\
 & \beta(\tau) & =  6c +\frac{{\check b}}{1+\tau}
+ \frac{2(2+\tau){\check W}}{(1+\tau)^2} + 4{\check W}'
+ \frac{3(1-\tau)(13+6\tau+\tau^2)}{4(6+\tau)(1+\tau)^3} +
\frac{2(17+3\tau)}{6+\tau}\Gamma(\tau) \nnb\\
& - & \frac{4(5-6\tau-\tau^2)}{6+\tau}\Gamma '(\tau) -
\frac{8\tau}{6+\tau}\Gamma ''(\tau) \nnb \\
 & \gamma(\tau) & = c  +\frac{{\check b}}{1+\tau}  +
\frac{(3+\tau){\check W}}{(1+\tau)^2} + \Gamma(\tau) \,.\eeqa
\noindent We now use the up to now free parameter ${\check b}$ to
reabsorb the parameter $c\,.$ Let us define
$$\bar{b} = {\check b} -
2c\,,\qq
\bar{W}(\tau) = {\check W}(\tau) + c(1+\tau)\,,$$
we get $$\widetilde{G}_{ij} = \bar{G}_{ij} + \bar{b} G_{ij} + \dst\mathop{\cal L}_{\vec{\bar W}}\,G_{ij}$$ 
with $\vec{\bar W} =\bar{W}(\tau){\vec{\phi}}$ and 
$$\dst \bar{G}_{ij} =
\widetilde{G}_{ij}\mid_{equ.(\ref{resultat})\ {\rm for}\ c = {\check
b} = {\check W}(\tau) \equiv 0} \,.$$

\vspace{1cm}
{\bf The dualised
$\mathbf{ SU(2)}\ \sigma$-model is therefore renormalisable at the two-loop order
if and only if we add a finite $\mathbf{\hbar}$ deformation of the classical
metric, depending on  two new parameters $\mathbf{\la_0^o}$ and
$\mathbf{\la_1^o}$ .}

\section{Concluding remarks}

We have been able to exhibit some set of counter-terms that ensures
the two-loop renormalisability of the T-dualised chiral non-linear
$\si$ model. The one-loop effective metric is defined {\bf up to two
constants} ($\la_0^o$ and
$\la_1^o$), and some finite arbitrary field and coupling
renormalisations. As is well known (e.g. in \cite{hull}), the two-loop
Callan-Symanzik
$\beta$ function (related to
$\Lambda_2$ \footnote{\,The two loops quantities $\Lambda_2$ and $\vec{W}_2$
are fixed as :
$$\Lambda_2 = \frac{\bar{b}}{2}\,,\qq \vec{W}_{2}\ {\rm obtained\ through\
(\ref{substitution})}\,.$$ Notice that the normalisation condition
$\bar{b} = 0$ (no
$\hbar$ extra finite coupling constant renormalisation) enforces
$\Lambda_2 = 0\,.$}) depends on these finite counterterms.

We emphasize that, contrarily to D. Friedan's approach to $\sigma$
models quantisation, where the classical metric receives {\bf infinite}
perturbative deformations, our candidate for the deformation of the
classical metric is a {\bf finite one}, depending on {\bf only two
parameters} (plus the usual infinite, and finite, renormalisations of the
fields and of the coupling constant) : our ansatz is that a proper
understanding of the dualisation process will precisely offer the extra
constraints that uniquely define the quantum extension of the classical
theory, order by order in perturbation theory, in the same spirit as Ward
identities determine what otherwise would appear as new parameters (see
also footnote 2).

\vspace{1cm}

\noindent{\bf Acknowledgments :} It is a pleasure to thank Galliano Valent
whose interest in that subject was really stimulating for us.


\begin{thebibliography}{99}

\bibitem{aab1} E. Alvarez, L. Alvarez-Gaum\' e, J. L. F. Barb\' on and Y.
Lozano,  {\sl Nucl. Phys.} {\bf B415} (1994) 71, {\tt hep-th/9309039}.


\bibitem{aal3} E. Alvarez, L. Alvarez-Gaum\' e and Y. Lozano, {\sl Nucl.
Phys.  Proc. Suppl.} {\bf 41} (1995) 1, {\tt hep-th/9410237}.


\bibitem{gpr} A. Giveon, M. Porrati and E. Rabinovici, {\sl Phys. Rep. }
{\bf 244} (1994) 77, {\tt hep-th/9401139}.

\bibitem{Al1} O. Alvarez, {\sl Nucl. Phys.} {\bf  B584} (2000) 659 ;
ib. 682, {\tt hep-th/0003177/0003178}.


\bibitem{grav}  A. Giveon, E. Rabinovici and G. Veneziano, {\sl Nucl.
Phys.} {\bf B322} (1989) 167.

\bibitem{CZ} T. Curtright and C. Zachos, {\sl Phys. Rev.} {\bf D49}
(1994) 5408 , {\tt hep-th/9401006}.

\bibitem{aal2} E. Alvarez, L. Alvarez-Gaum\' e and Y. Lozano,  {\sl
Phys. Lett.}  {\bf B336} (1994) 183, {\tt hep-th/9406206}.

\bibitem{Lo} Y. Lozano, {\sl Phys. Lett.} {\bf B355} (1995) 165,  {\tt
hep-th/9503045}.

\bibitem{Sf} K. Sfetsos, {\sl Phys. Rev.} {\bf D54} (1996) 1682,  {\tt
hep-th/9602179}.

\bibitem{fj} B. E. Fridling and A. Jevicki, {\sl Phys. Lett.} {\bf B134}
(1984) 70.
\bibitem{ft} E. S. Fradkin and A. A. Tseytlin, {\sl Ann. Phys.} {\bf 162}
(1985) 31.

\bibitem{cv} P.Y. Casteill and G. Valent, {\sl Nucl. Phys.} {\bf B591}
(2000) 491	,  {\tt hep-th/0006186}.

\bibitem{st} A. Subbotin and I.V. Tyutin,  {\sl Int. J. Mod. Phys.} {\bf A11} (1996) 1315  {\tt
hep-th/9506132}.

\bibitem{kt} J. Balog, P. Forg\` acs, Z. Horv\'ath, L. Palla,  {\sl
Nucl. Phys. Proc. Suppl.} {\bf 49} (1996) 16,  {\tt
hep-th/9601091}.

\bibitem{kt2} J. Balog, P. Forg\` acs, Z. Horv\'ath, L. Palla,  {\sl
Phys. Lett.} {\bf B388} (1996) 121,  {\tt
hep-th/9606187}.

\bibitem{km} N. Kaloper and K. Meissner,  {\sl Phys. Rev.} {\bf D56} (1997) 7940,  {\tt hep-th/9705193}.

\bibitem {bd} G. Bonneau and F. Delduc, {\sl Nucl. Phys.} {\bf B250} (1985) 581.

\bibitem{bo} G. Bonneau, {\sl Int. J. Mod. Phys.} {\bf A5} (1990) 3831.

\bibitem{Fr} D. Friedan, {\sl Ann. Phys.} {\bf 163} (1985) 1257.

\bibitem{HT} C.M. Hull and P.K. Townsend, {\sl Phys. Lett.} {\bf B191}
(1987) 115.

\bibitem{mt} R.R. Metsaev and A.A. Tseytlin, {\sl Phys. Lett.} {\bf B191}
(1987) 354.

\bibitem{z} D. Zanon, {\sl Phys. Lett.} {\bf B191} (1987) 363.

\bibitem{Ince} E.L. Ince, \textsl{Ordinary differential equations} (Dover Publications, New York, 1956), chapter 16.

\bibitem{hull} C.M. Hull, {\sl ``Lectures on non-linear sigma models and
strings"}, 1986 Workshop ``Super Field Theories", Vancouver, Canada ;
published in Vancouver Theory Worshop 1986:77, {\tt Cambridge
Preprint-87-0480}.



\end{thebibliography}
\end{document}